\documentclass[twocolumn,preprintnumbers,amsmath,amssymb]{revtex4}
\usepackage{graphicx} 
\usepackage{setspace} 
\usepackage{dcolumn} 
\usepackage{bm} 
\usepackage{natbib}
\usepackage{hyperref}
\hypersetup{colorlinks=true,linkcolor=blue,citecolor=blue,urlcolor=blue}
\usepackage{epsfig}
\usepackage{amsmath}
\begin{document}
\title{Structural and Thermoelectric Properties of Al$_{1+x}$Fe$_{2-y}$M$_y$B$_2$ (M=Ag, Ni, Sb, Ga, and Ge) Intermetallic Borides}
\author{D. Sivaprahasam$^a$\footnote{Email: sprakash@arci.res.ac.in}}
\author{R. Preyadarshini$^a$}
\author{Ashutosh Kumar$^b$}

\affiliation{$^a$Centre for Automotive Energy Materials (CAEM), International Advanced Research Centre
for Powder Metallurgy and New Materials (ARCI), IITM Research Park, Taramani, Chennai – 600 113, India}
\affiliation{$^b$Functional Materials Laboratory, Department of Materials Science and Metallurgical Engineering, Indian Institute of Technology Bhilai - 491 002, India}
\date{\today}
\begin{abstract}
AlFe$_2$B$_2$, a ternary transition metal boride doped with Ag, Ni, Sb, Ga, and Ge was investigated for the phase constituents, microstructure, and thermoelectric properties. The parent compound with 20\% excess Al (Al$_{1.2}$Fe$_2$B$_2$) prepared by vacuum arc melting contains orthorhombic AlFe$_2$B$_2$ and FeB upon doping form additional phases such as Ag$_5$Al, AlNi$_3$, Al$_3$Ni$_2$, AlSb, and AlB$_2$. The structural parameters of the AlFe$_2$B$_2$ phase in Ag, Ni, and Sb-doped samples are nearly the same as the un-doped one. The lattice parameters and cell volume increase noticeably in Ga and Ge-doped compounds. The Ga-doped sample shows only FeB as a secondary phase; however, the AlB$_2$ phase is observed in Ge-doped AlFe$_2$B$_2$.  The microstructure investigated in FEG-SEM with EDS shows the AlFe$_2$B$_2$ matrix is chemically homogenous in all samples, except the Ga-doped one, with uniformly distributed single or multiple secondary phases. The DSC-TG results show that all the dopants decrease the AlFe$_2$B$_2$ phase peritectic decomposition reaction temperature, indicating the structure destabilizes. The thermoelectric properties measured show that AlFe$_2$B$_2$ is an n-type compound with electrical conductivity in the range of 0.35-0.46$\times$10$^6$ S/m. The addition of Ag, Ni, Ga, and Ge marginally alters the Seebeck and electrical conductivity. The noticeable improvement in Seebeck coefficient and with negligible change in electrical conductivity resulted in a highest power factor of 0.4 mW/mK$^2$ for Sb-doped sample..  The thermal conductivity of AlFe$_2$B$_2$, which is in the range of 6.4-9.1 W/m·K between 300 to 773\,K, decreases with Sb doping to 5.2-8.5 W/m·K resulting in a maximum zT of 0.04 at 773\,K.                 
\end{abstract}
\maketitle
\section{Introduction}
In the relentless pursuit of sustainable energy solutions, the quest for highly efficient and versatile materials has led researchers to the forefront of thermoelectric (TE) innovation. As we stand on the cusp of a new era in energy technologies, the emergence of novel TE materials promises to revolutionize how we harness and utilize heat as a valuable resource.\cite{R1} Over a long period, TE materials have faced inherent limitations of low efficiency and high cost. However, breakthroughs in materials science in recent decades have given rise to a new generation of TE materials that exhibit unprecedented performance, opening up new possibilities for widespread applications like power from industrial waste heat, automotive exhaust, solar thermal, and body heat for wearable electronics/sensors. The performance of a TE material is evaluated by a dimensionless figure of merit, zT=S$^2\sigma$T/$\kappa$, where S is the Seebeck coefficient ($\mu$V/K), $\sigma$-electrical conductivity (S/m), $\kappa$-thermal conductivity (W/m·K), and T is the absolute temperature (K). Although optimization of these TE parameters is complex due to their interdependence, several innovative strategies have led to developing TE materials with zT higher than 1, in a few compounds zT$>$2 are also achieved.\cite{R1} However, the fabrication of TE devices from these materials for the medium- and high-temperature range is not encouraged widely due to their high cost, and hence high-performance, low-cost earth-abundant materials are solicited.\\ 
A strong interest in boron$-$based compounds for thermoelectric has been sparked by the finding of very high Seebeck coefficients, S(T) $\sim$ 250 $\mu$V/K, at high carrier densities (n$\sim$ 10$^{21}$ cm$^{-3}$).\cite{R2}
Of all the known boron-containing TE materials, intermetallic borides are particularly interesting because of their enormous potential in TE generators, which can operate in extremely harsh environments (T$>$1000\,K).\cite{R2,R3} AlFe$_2$B$_2$ is a promising magneto-calorific material garnering attention as a potential TE material due to its unique properties and cost-effective composition.\cite{R4,R5,R6,R7} The orthorhombic ($Cmmm$) structured intermetallic compound exhibits a metallic band structure that maintains a covalent-ionic character.\cite{R8,R9} The Fe atoms are bonded to four nearby $B$ atoms at a distance of 2.156$\AA$ and two further $B$ at a distance of 2.165$\AA$, forming a slab-type layer, and a layer of Al atoms separates these layers.\cite{R10} The synthesis of AlFe$_2$B$_2$ typically involves methods, such as arc melting and reactive hot pressing.\cite{R10,R11,R12} Melting, though popular, the peritectic reaction (S$_1+$Liquid S$_2$) during solidification has always resulted in a FeB phase in Al$\colon$Fe$\colon$B (1$\colon$2$\colon$2), which has high electrical and thermal conductivities.\cite{R13} Adding excess Al avoids FeB, but it could form an Al$_{13}$Fe$_4$ phase beyond a limit. Treating in dilute HCl leaches out the Al$_{13}$Fe$_4$ phase, making the AlFe$_2$B$_2$ more pure. However, this method can’t remove the FeB phase. The ferromagnetic FeB is deleterious to magneto-caloric application. Nevertheless, for TE use, $\sigma$ and $\kappa$ of the secondary phase are the critical properties. The electrical conductivity of FeB and Al$_{13}$Fe$_4$ phases is comparable. However, the thermal conductivity of Al$_{13}$Fe$_4$ is significantly higher than FeB and AlFe$_2$B$_2$.\cite{R13,R14} Hence, depending on the synthesis method and conditions, Al in the overall initial stoichiometry must be fixed carefully to prevent the formation of Al$_{13}$Fe$_4$ and minimize the FeB phase content in the end product.\\      
This work investigates the structure, microstructure, and thermoelectric properties of Al$_{1.2}$Fe$_2$B$_2$ doped with various elements such as Ag, Ni, Sb, Ge, and Ga. Feedstock taken at the required stoichiometry were arc melted, annealed, powdered, vacuum hot-pressed to pellets, and investigated for their phase content, microstructure, thermo-physical, and thermoelectric properties. The compound exhibits n-type conduction with high residual electrical conductivity and thermal conductivity. In most cases, doping forms an additional secondary phase/s apart from FeB. In Sb addition alone, the transport properties are altered, resulting in an improvement in zT.\\
\section{Experimental Section}
The AlFe$_{2-x}$M$_x$B$_2$ compounds were synthesized by vacuum arc melting of Al, Fe$_2$B, B, and M (Ag, Ni, Sb, Ge, Ga) pieces taken in the required stoichiometry. All constituents' purity was 99.9\% except B, which is 99.5\% pure. Before melting, the chamber was evacuated to below 5 $\times$10$^{-5}$ mbar vacuum and purged with dynamic argon flow at 300 mbar. Each compound’s sample weighing 10g was melted three times after flipping every melting. The samples thus prepared were annealed at 1323\,K for 96 h after sealing in a quartz tube under a 5$\times$10$^{-5}$ mbar vacuum to improve their chemical homogeneity. The annealed samples were subsequently crushed and ground to fine powder, consolidated into 20 mm diameter pellets in a vacuum hot press (Thermal Technology, USA) at 1273\,K for 2\,h under 50 MPa pressure. The hot pressing was carried out in a graphite die under the 5-8$\times$10$^{-5}$ mbar vacuum. The constituent phases in the synthesized powder and the pellets were characterized using a Rigaku Smartlab X-ray diffractometer (Cu-K$_{\alpha}$, $\lambda$=0.15406\,nm) with a step of 0.01$^{\circ}$ and scan range from 20-80$^{\circ}$ at 1$^{\circ}$/min scan rate. The lattice parameters ‘$a$’, ‘$b$’, and ‘$c$’ of the orthorhombic AlFe$_2$B$_2$ phase were calculated from d$_{(200)}$, d$_{(060)}$, and d$_{(001)}$ diffraction peaks. The sample’s microstructure was analyzed in FE-SEM (Zeiss Merlin, Germany), and chemical analysis of various phases present in the microstructure was carried out using energy-dispersive X-ray spectrometry (EDS). The thermal characteristics were evaluated using differential scanning calorimetry-thermogravimetry (DSC-TG) in a Setaram Instrumentation, Lab Sys Evo setup. The Seebeck coefficient and electrical conductivity were measured using a standard four-probe method (Seebsys system: NorECS AS, Norway). The spatial distribution of the Seeback coefficient at room temperature was measured using a potential Seeback microprobe (PSM). The thermal conductivity of the sintered samples from room temperature to 773\,K was measured using a transient plane source technique (Hot Disk TPS 2500, Sweden). The sample's thermal conductivities ($\kappa$=D$\times$C$_p\times\rho$) were also measured using the thermal diffusivity (D), specific heat (C$_p$), and sample density ($\rho$). The D measurements were performed using the laser flash technique (LFA–457, Netsch GmbH, Germany) on a cylindrical sample having a 10\,mm diameter and $\sim$2\,mm thickness. The C$_p$ was obtained using Dulong–Petit law. The density ($\rho$) of the hot-pressed pellet was measured using the Archimedes method.\\
\begin{table*}
\caption{Lattice parameter and secondary phases observed in Al$_{1+x}$Fe$_2$B$_2$ and Al$_{1.2}$Fe$_{0.9}$M$_{0.1}$B$_2$.}
\centering
\begin{tabular}{c c c c c c c}
\hline
sample & Phase & &Lattice parameters& & Secondary & Volume\\
 & & a ($\AA$) & b ($\AA$) & c ($\AA$) & Phase & ($\AA^3$)\\
\hline
\hline
Al$_{1.1}$Fe$_2$B$_2$ & AlFe$_2$B$_2$ & 2.907(5) & 11.000(3) & 2.866(2) &FeB, Fe$_2$B& 91.67\\
\hline
Al$_{1.2}$Fe$_2$B$_2$ & AlFe$_2$B$_2$ & 2.924(5) & 11.003(3) & 2.871(1) &FeB& 92.61\\
\hline
Al$_{1+x}$Fe$_2$B$_2$ & AlFe$_2$B$_2$$^*$\cite{R9} & 2.916(8) & 11.033(2) & 2.866(0) & & 92.23\\
 & AlFe$_2$B$_2$\cite{R15} & 2.907(5) & 11.000(3) & 2.866(2) & & 92.23\\
 & AlFe$_2$B$_2$\cite{R5} & 2.924(1) & 11.033(9) & 2.870(1) & & 92.60\\
 & AlFe$_2$B$_2$\cite{R18} & 2.924(1) & 11.029(9) & 2.866(0) & & \\
 & AlFe$_2$B$_2$\cite{R19} & 2.916(2) & 11.022(5) & 2.851(5) & & 91.66 \\
 & AlFe$_2$B$_2$\cite{R20} & 2.926(3) & 11.029(5) & 2.866(6) & FeB& 92.52\\
 & AlFe$_2$B$_2$\cite{R21} & 2.925(1) & 11.029(6) & 2.868(5) & & \\
 & AlFe$_2$B$_2$ & 2.931(1) & 11.028(4) & 2.861(1) & & 92.50\\
 \hline
 Al$_{1.2}$Ag$_{0.1}$Fe$_{1.9}$B$_2$ & AlFe$_2$B$_2$ & 2.923(0) & 11.006(5) & 2.861(4) &FeB, Ag$_5$Al & 92.05\\
 \hline
 Al$_{1.2}$Ni$_{0.1}$Fe$_{1.9}$B$_2$ & AlFe$_2$B$_2$ & 2.925(8) & 11.020(3) & 2.861(1) & FeB, AlNi$_3$, Al$_3$Ni$_2$ & 92.25\\
 \hline
 Al$_{1.2}$Ni$_{0.2}$Fe$_{1.8}$B$_2$ & AlFe$_2$B$_2$\cite{R22} & 2.927(1) & 11.041(3) & 2.869(1) & & 92.71\\
 & AlNi$_2$B$_2$\cite{R18} & 2.977(9) & 11.040(3) & 2.849(7) & & 93.68\\ 
\hline
AlSb$_{0.1}$Fe$_{1.9}$B$_2$& AlFe$_2$B$_2$ & 2.915(4) & 11.025(8) & 2.873(7) &FeB, Al$_{13}$Fe$_4$ & 92.37\\
\hline
AlGa$_{0.1}$Fe$_{1.9}$B$_2$& AlFe$_2$B$_2$ & 2.875(4) & 11.055(9) & 2.880(8) &FeB, Al$_{13}$Fe$_4$ & 91.58\\
& Al$_{1.1}$Ga$_{0.1}$Fe$_{1.9}$B$_2$\cite{R23} & 2.927(2) & 11.035(1) & 2.870(2) & & 92.72\\
\hline
AlGe$_{0.1}$Fe$_{1.9}$B$_2$& AlFe$_2$B$_2$ & 2.933(7) & 11.054(2) & 2.877(8) &AlB$_2$& 93.33\\
& Al$_{1.1}$Ge$_{0.1}$Fe$_2$B$_2$\cite{R23} & 2.928(4) & 11.036(9) & 2.870(0) & & 92.75\\
\hline
$*$-Single crystal
\end{tabular}
\label{Table I}
\end{table*}
\section{Results and Discussion} 
\subsection{Phases and Crystal Structure}
\begin{figure*}
\centering
  \includegraphics[width=0.7\linewidth]{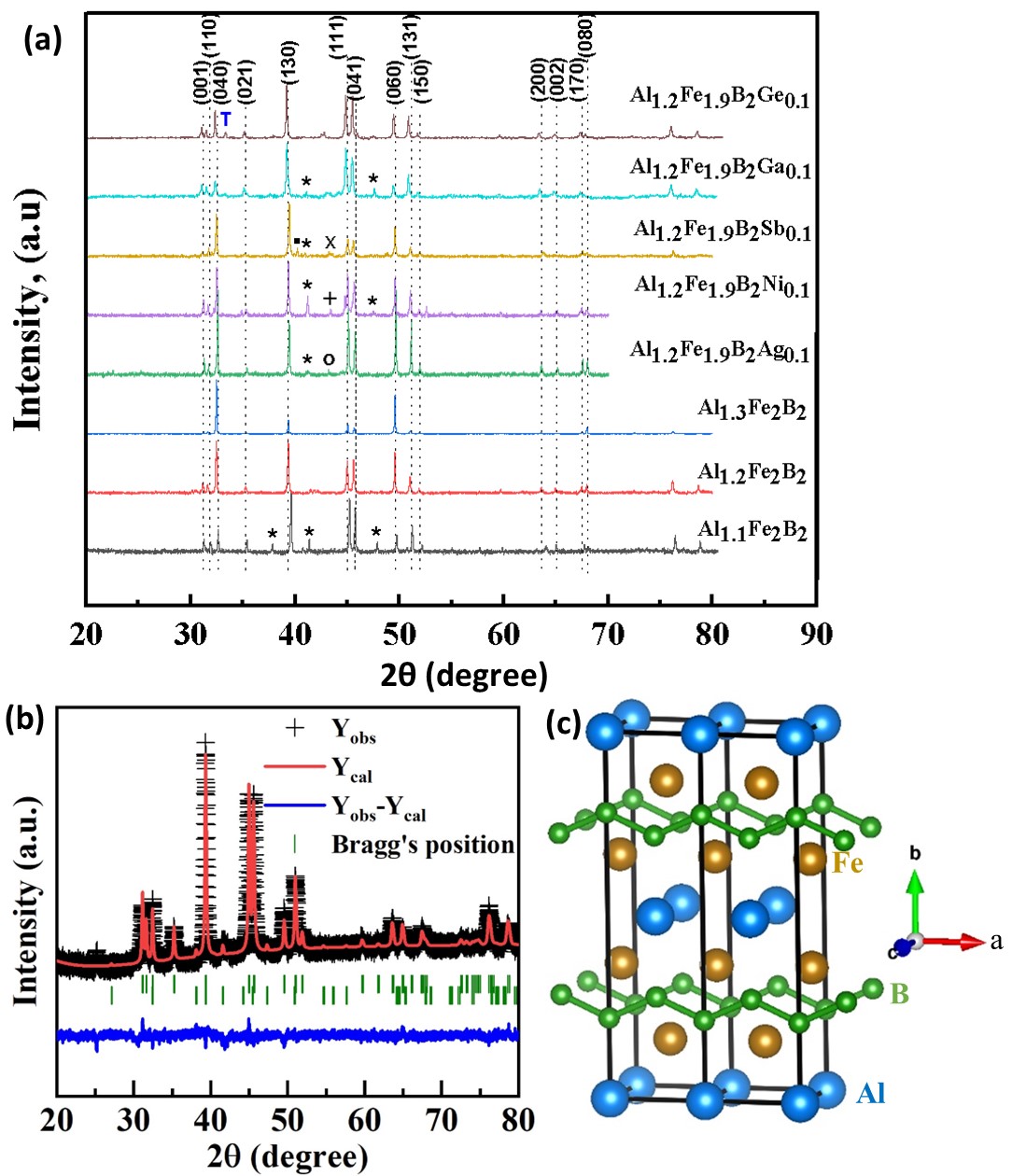}
  \caption{XRD Pattern of Al$_{1+x}$Fe$_2$B$_2$ and Al$_{1.2}$Fe$_{2-x}$M$_x$B$_2$ (M=Ag, Ni, Sb, Ge, and Ga). ($*$)-FeB, ($+$)-AlNi$_3$, ($o$)-Ag$_5$Al, ($T$)-AlB$_2$ and ($.$)- Unknown. (b) Rietveld refinement pattern for Al$_{1.2}$Fe$_2$B$_2$ and (c) corresponding crystal structure using Vesta software.}
  \label{fig1a}
\end{figure*}

The peritectic reaction during solidification of the AlFe$_2$B$_2$ melt demands careful assessment of excess Al addition required to obtain the minimal impurity (FeB or Al$_{13}$Fe$_4$) phases. Reports available indicate excess Al ranging from 20-200\% have been used to avoid the FeB; however, 30-50\% was observed to be ideal.\cite{R15,R16,R17} In the present work, since we want to avoid the Al$_{13}$Fe$_4$ phase and minimize the FeB in the end product, 10-30\% of Al addition over the stoichiometry was tested. The XRD pattern of the Al$_{1+x}$Fe$_2$B$_2$ ($x$= 0.1, 0.2, and 0.3) along with the optimized stoichiometry Al$_{1.2}$Fe$_2$B$_2$ contain different dopants Al$_{1.2}$Fe$_{1.9}$M$_{0.1}$B$_2$, (M= Ag, Ni, Sb, Ga, and Ge) are given in Fig.\ref{fig1a}(a). The Rietveld refinement of Al$_{1.2}$Fe$_2$B$_2$ is shown in Fig.\ref{fig1a}(b) and corresponding crystal structure obtained using the refined structural parameters is shown in Fig.\ref{fig1a}(c).\\
The 10\% excess Al sample (Al$_{1.1}$Fe$_2$B$_2$) XRD pattern shows a significant fraction of FeB (98-009-2223) and Fe$_2$B (98-011-3520) along with AlFe$_2$B$_2$ (98-000-7593) phase. The FeB decreases to 6.9\% in Al$_{1.2}$Fe$_2$B$_2$ and is free from Fe$_2$B in 20\% excess Al sample. The FeB further reduces to below 1\% in Al$_{1.3}$Fe$_2$B$_2$. The dopant-added samples were prepared by substituting Fe with 0.1 part in Al$_{1.2}$Fe$_2$B$_2$ stoichiometry. Though Al$_{1.3}$Fe$_2$B$_2$ gives relatively lesser FeB, it results in a high volume fraction of other (Al$_x$M$_y$) secondary phases, and hence, the Al$_{1.2}$Fe$_2$B$_2$ composition was chosen for doping.  Further, as stated earlier, the FeB has relatively lesser thermal conductivity than the Al$_{13}$Fe$_4$ and starts appearing with higher Al, which harms TE properties.\cite{R9} The lattice parameters, unit cell volume, and the secondary phases present in Al$_{1.2}$Fe$_2$B$_2$ and doped Al$_{1.2}$Fe$_{1.9}$M$_{0.1}$B$_2$ samples, along with other reported data are listed in Table~\ref{Table I}.\\
The lattice parameters and the cell volume estimated from the XRD pattern of un-doped Al$_{1.2}$Fe$_2$B$_2$ are comparable to the other reported values for Al$_{l+x}$Fe$_2$B$_2$ samples with secondary phase/s.\cite{R15,R22} The Ag, Ni, and Sb doping decreases the lattice parameter marginally, and Ag doping shows the maximum deviation among these three. The unit cell volume corresponding to the Ag, Ni, and Sb doped samples decreases by 0.6\%, 0.38\%, and 0.21\%, respectively. In the Ni-doped AlFe$_2$B$_2$, Lejeune et al. observed that the lattice parameter and the cell volume increase with Ni content in the compound.\cite{R22} The relatively lesser cell volume observed compared to the un-doped Al$_{1.2}$Fe$_2$B$_2$ in the present study possibly arises from the Al depletion in AlFe$_2$B$_2$ crystal due to the formation of AlNi$_3$ and Al$_3$Ni$_2$ phases. The Ga-doped compound shows lattice parameters and cell volume higher than the undoped AlFe$_2$B$_2$. In the Ge-doped compound, the lattice parameter increases, particularly in the “$a$” and “$c$” directions resulting in 1.4\% higher cell volume compared to pristine AlFe$_2$B$_2$.\\   
\begin{figure*}
\centering
  \includegraphics[width=0.8\linewidth]{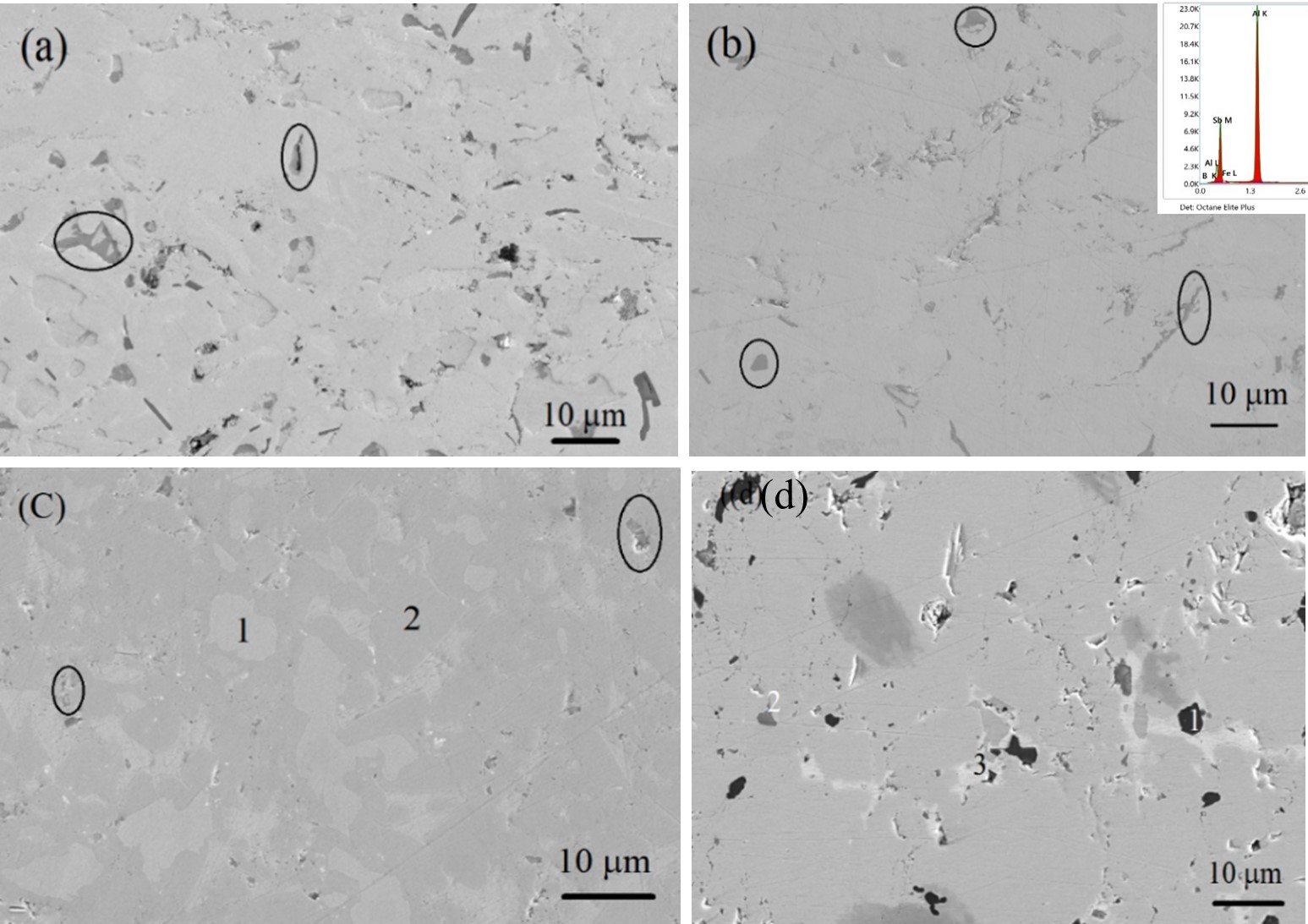}
  \caption{SEM micrographs of (a) Al$_{1.2}$Fe$_2$B$_2$ (b) Al$_{1.2}$Fe$_{1.9}$Sb$_{0.1}$B$_2$ (c) Al$_{1.2}$Fe$_{1.9}$Ga$_{0.1}$B$_2$ and (d) Al$_{1.2}$Fe$_{1.9}$Ge$_{0.1}$B$_2$}.
  \label{fig2a}
\end{figure*}
The XRD pattern of Al$_{1.2}$Fe$_{1.9}$Ag$_{0.1}$B$_2$ and Al$_{1.2}$Fe$_{1.9}$Ni$_{0.1}$B$_2$ showed a secondary FeB phase. Additionally, the Ag-doped compound contains a small fraction of the Ag$_5$Al phase, and the Ni-doped sample shows the presence of AlNi$_3$ and Al$_3$Ni$_2$ phases. The XRD diffraction pattern of the Sb substituted sample also shows a similar diffraction pattern, with additional impurity peaks corresponding to Al$_{13}$Fe$_4$, along with the AlFe$_2$B$_2$ phase. However, the FeB phase quantified by Rietveld refinement is significantly low (0.58\%). This indicates that the Sb doping reduces the FeB content in the solidified sample. The Al$_{1.2}$Fe$_{1.9}$Ga$_{0.1}$B$_2$ compound has a significant fraction of FeB and Al$_{13}$Fe$_4$; however, the Al$_{1.2}$Fe$_{1.9}$Ge$_{0.1}$B$_2$ shows no diffraction peak corresponding to FeB, and only AlB$_2$ secondary phase observed. Experimental work reported by Barua et al. suggests that Ga and Ge have minimal solid solubility in AlFe$_2$B$_2$.\cite{R23} The Ga and Ge doping used here is within the range of the solid solubility. The formation of a higher fraction of secondary phases can be attributed to the enhanced Al, Fe, and B vacancies, as suggested by Samanta et al. near the dopant atoms.\cite{R24}\\
\subsection{Microstructure}
The microstructural features of the Al$_{1.2}$Fe$_2$B$_2$ and Al$_{1.2}$Fe$_{1.9}$M$_{0.1}$B$_2$ (M=Sb, Ga, and Ge) samples were investigated after polishing the cross-section to 0.2 $\mu$m finish using diamond paste. Fig.~\ref{fig2a}(a-d) shows the SEM micrographs of the Al$_{1.2}$Fe$_2$B$_2$ and the Al$_{1.2}$Fe$_{1.9}$M$_{0.1}$B$_2$ (M=Sb, Ga, Ge) taken in secondary electron mode. The Al$_{1.2}$Fe$_2$B$_2$ microstructure consists of a matrix containing Al, Fe, and B with the secondary phase of an irregular morphology. The EDS analysis of the secondary phase (encircled in Figure~\ref{fig2a}a) showed only Fe and B, possibly the FeB phase.\\ 
The Al$_{1.2}$Fe$_{1.9}$M$_x$B$_2$ sample microstructures are also similar to Al$_{1.2}$Fe$_2$B$_2$. In all cases, the microstructure shows multiple secondary phases in the AlFe$_2$B$_2$ matrix. For example, in the Al$_{1.2}$Fe$_{1.9}$Sb$_{0.1}$B$_2$ sample, along with FeB, another secondary phase containing only Al and Sb is present in the AlFe$_2$B$_2$ matrix (Figure~\ref{fig2a}b). The volume fraction of FeB is relatively lesser than the undoped Al$_{1.2}$Fe$_2$B$_2$ sample. The quantitative compositional analysis of the secondary phase showing Al and Sb shows notable variation in the Al/Sb ratio, with most of the particles being close to 1, indicating these particles could be AlSb intermetallic phase. Though the XRD of Al$_{1.2}$Fe$_{1.9}$Sb$_{0.1}$B$_2$ does not show any diffraction peak corresponding to this phase, the sample prepared with higher Sb doping (Al$_{1.2}$Fe$_{1.8}$Sb$_{0.2}$B$_2$) showed prominent peaks corresponding to the AlSb phase (Figure~S1) and the microstructure also has a higher volume of these particles. The Ga-doped samples (Fig.~\ref{fig2a}c) showed the AlFe$_2$B$_2$ matrix phase with two types of grains (marked 1 and 2 in Fig.~\ref{fig2a}c) predominantly consisting of Al, Fe, and B with different ratios and the FeB secondary phase. Few particles containing Al, and Fe with high percentage of Ga were also observed.  In the Ge-doped sample, the microstructure (Fig.~\ref{fig2a}d) consists of multiple secondary phases in the AlFe$_2$B$_2$ matrix. The matrix is relatively inhomogeneous on the micro-scale compared to other doped samples. Secondary phase/s of composition AlxBy with the high atomic percentage of B (marked as 1, 2 in Fig.~\ref{fig2a}d) have been observed throughout the microstructure. The Ge in the matrix is below the detectable limit of EDS. However, some white regions (marked as 3) contain all four elements B, Fe, Al, and Ge, indicating Ge dissolved in the AlFe$_2$B$_2$ matrix.\\
The structural details and the microstructure analysis of the Al$_{1.2}$Fe$_2$B$_2$ and AlFe$_{1.9}$M$_{0.1}$B$_2$ suggest that the Al addition over the required 1:2:2 stoichiometry plays a crucial role in the type and volume of the secondary phase/s generated. The dopants added were 2 at.\% in the overall stoichiometry, predominantly formed secondary phase/s with Al. Among the dopants examined, Ni exhibits the highest tendency to make secondary phases with Al form AlNi$_3$ (98-006-5926) and Al$_3$Ni$_2$ (98-009-8815) phases. Such phases deplete the excess Al available, resulting in the production of a high percentage of FeB. Kadas et al. suggest that Ni substitution at the Fe site in AlFe$_2$B$_2$ is energetically not favorable.\cite{R18} The lattice parameters of the AlFe$_{1.9}$Ni$_x$B$_2$ sample are almost the same as AlFe$_2$B$_2$, supporting the above fact. The Ag doping also showed a similar effect that produces the FeB and Ag$_5$Al phases. The most significant effect of doping in the microstructure was observed in the Al$_{1.2}$Fe$_{1.9}$Ge$_{0.1}$B$_2$ sample, where the FeB phase is insignificant, though a small amount of other unknown impurities were observed. The Sb-doped AlFe$_2$B$_2$ also showed a very low FeB phase.\\
\subsection{DSC-TG Analysis}
\begin{figure}
\centering
  \includegraphics[width=0.99\linewidth]{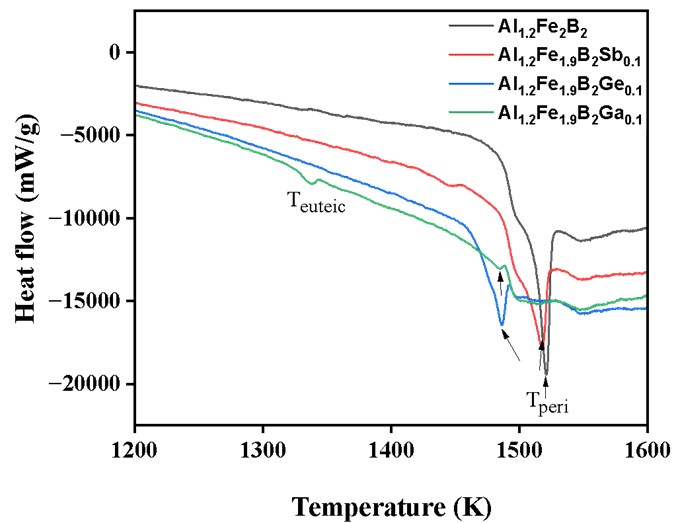}
  \caption{DSC curve of Al$_{1.2}$Fe$_2$B$_2$ and Al$_{1.2}$Fe$_{1.9}$M$_{0.1}$B$_2$ (M=Sb, Ga, Ge) compounds. The Tperi temperature is arrow-marked.}
  \label{fig3a}
\end{figure}

The Al$_{1.2}$Fe$_2$B$_2$ and Al$_{1.2}$Fe$_{1.9}$M$_{0.1}$B$_2$ (M=Sb, Ga, Ge) samples were characterized for their thermal attributes using DSC-TG in the temperature range 300\,K $<$ T $<$ 1673\,K are shown in Fig.~\ref{fig3a}. Table~\ref{Table II} lists the endo and exo-thermic peaks observed in the DSC-TG test while heating and cooling of Al$_{1.2}$Fe$_2$B$_2$ and Al$_{1.2}$Fe$_{1.9}$M$_{0.1}$B$_2$.\\
\begin{table*}
\caption{Temperatures corresponding to melting (T$_m$), peritectic (T$_{peri}$) and eutectic (T$_{eutec.}$) reactions.}
\centering
\begin{tabular}{c c c c c c}
\hline
Compound & T$_m$ (K) & T$_{peri}$ (K) & & T$_{eutec.}$ (K) &\\
         & & Heating & Cooling & Heating & Cooling \\
\hline
\hline
Al$_{1.2}$Fe$_2$B$_2$ & 1620 & 1521 & 1491 & & 1313\\
\hline
Al$_{1.2}$Sb$_{0.1}$Fe$_{1.9}$B$_2$ & 1581 & 1517 & 1491 & 1445& 1341\\
\hline
Al$_{1.2}$Ge$_{0.1}$Fe$_{1.9}$B$_2$ & 1617 & 1485 & 1455 & & \\
\hline
Al$_{1.2}$Ga$_{0.1}$Fe$_{1.9}$B$_2$ & 1535 & 1482 & 1455 & 1337& 1251\\
\hline
\hline
\end{tabular}
\label{Table II}
\end{table*}
The liquidus temperature of the Al$_{1.2}$Fe$_2$B$_2$ obtained from the cooling curve (Fig.\ref{fig3a}) is comparable with the other reported value for similar stoichiometry.\cite{R6} While Ge doping decreases the Tm marginally, Sb and Ga doping reduces the T$_m$ to a notable extent.  In the Al$_{1.2}$Fe$_2$B$_2$ sample, an endothermic peak with an onset and peak temperature of 1507\,K and 1521\,K, respectively is associated with the peritectic decomposition (T$_{peri}$) reaction of the AlFe$_2$B$_2$ into FeB and the corresponding liquid upon heating. In the pure Al$_{1.2}$Fe$_2$B$_2$ phase prepared by arc melting followed by suction casting, the endothermic peak corresponding to peritectic reaction is observed at 1551\,K.\cite{R5,R23} The presence of FeB secondary phase affects the T$_{peri}$, which is evident in Al$_{1.2}$Fe$_2$B$_2$ with 2\%FeB prepared and analyzed by Cedervell et al. showing T$_{peri}$ of 1515\,K.\cite{R19} In Sb-doped samples, the decomposition starts at 1491\,K and peaks at 1517\,K. For Ga and Ge doping, exothermic starts at 1462\,K and 1469\,K, respectively, is notably lesser than un-doped Al$_{1.2}$Fe$_2$B$_2$. Barua et. al., attribute such a decrease of T$_{peri}$ in Ga and Ge doping to the overall Al stoichiometry.\cite{R23} For example, the T$_p$ reaches a minimum value of 1511\,K in the sample of composition Al$_{1.2}$Fe$_{1.9}$Ga$_{0.1}$B$_2$. The lower T$_{peri}$ observed in this work in both un-doped and Sb, Ge, Ga doped Al$_{1.2}$Fe$_2$B$_2$ is a consequence of the partial substitution of Al and Fe in the Al$_{1.2}$Fe$_2$B$_2$ phase. The temperature differences between T$_m$ and T$_{peri}$ of doped compounds are significantly different from Al$_{1.2}$Fe$_2$B$_2$ under equilibrium cooling conditions. As per the equilibrium phase diagram, the formed volume fraction of FeB is determined by the gap between Tm and T$_{peri}$. The small percentage of FeB in the Sb-doped system compared to Al$_{1.2}$Fe$_2$B$_2$ may be attributed to the relatively lesser temperature gap between T$_m$ and T$_{peri}$.\\
\subsection{Thermoelectric Properties}
\begin{figure*}
    \centering
  \includegraphics[width=0.8\linewidth]{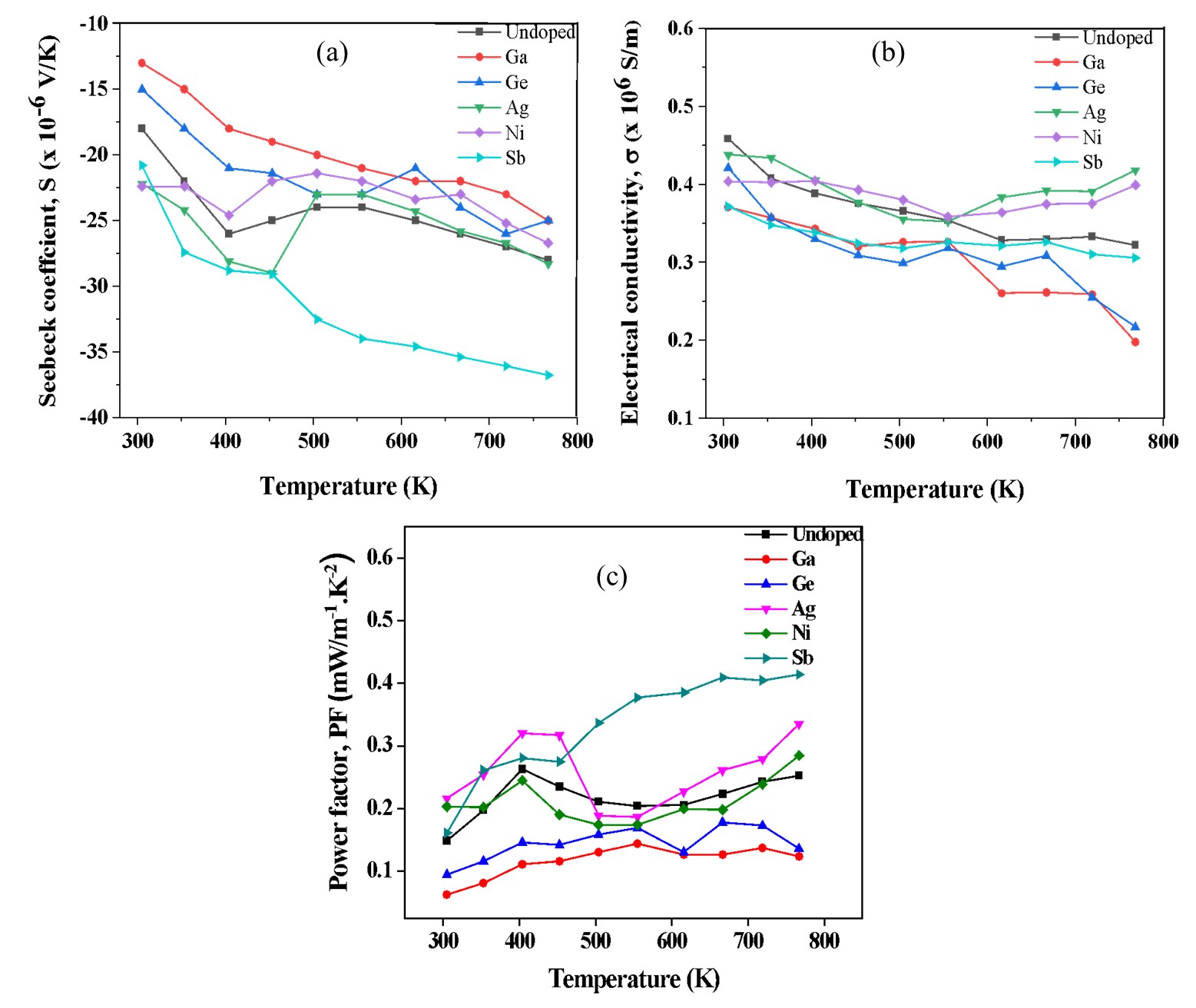}
  \caption{Seebeck coefficient, electrical conductivity, and powder factors of Al$_{1.2}$Fe$_2$B$_2$ and Al$_{1.2}$Fe$_{1.9}$M$_{0.1}$B$_2$ (M= Ag, Ni, Sb, Ga, and Ge) samples.}
  \label{fig4a}
\end{figure*}

The Seebeck coefficient, electrical conductivity, and power factor of Al$_{1.2}$Fe$_2$B$_2$ and Al$_{1.2}$Fe$_{1.9}$M$_{0.1}$B$_2$ compounds evaluated from room temperature to 773\,K are given in Fig.~\ref{fig4a}(a-c).\\
All the compounds show a negative Seebeck coefficient, indicating n-type conduction, and the absolute value of thermopower |S| increases with the temperature, typically like a degenerate semiconductor. The S vs. T behavior obtained is consistent with the previously reported result of Levin et al. in arc melted, suction cast AlFe$_2$B$_2$ where the S varies between 25-35$\mu$V/K in the temperature range of 300-673\,K.\cite{R5} Lejeune et al also report similar S values in the 300-373\,K range.\cite{R24} The AlFe$_2$B$_2$ prepared by the melt route in this work, and by others, invariably contain FeB as a secondary phase, though the volume percent can be different depending upon the processing condition. FeB, unlike AlFe$_2$B$_2$, exhibits p-type conduction, and the Seebeck coefficient varies from 8 to 32$\mu$V/K between 440-960\,K.\cite{R26} To gain deeper insight into how secondary phases influence the Seebeck coefficient, the spatial distribution of S mapped over a 5 mm $\times$ 5 mm area using potential Seebeck measurement (PSM) system at room temperature was done, as shown in Fig.~S3. This shows two regions of the secondary phases with S values, one lower and another higher than the matrix phase (Al$_{1.2}$Fe$_2$B$_2$). As seen in the PSM map, a relatively higher value of Seebeck coefficient is observed at several locations, which may be attributed to the presence of FeB phase (as can be seen from SEM images). A similar rise in the Seebeck coefficient values are observed in the GeTe-WC \cite{R27}, PbTe-CoSb$_3$ \cite{R28}, and some oxide composites as well \cite{R29}, where the presence of interface barrier due to different Fermi level positions result in an enhancement in the Seebeck coefficient. This indicates that the FeB secondary phases present, to a notable extent, increase the overall Seebeck coefficient. Hence, the intrinsic Seebeck coefficient of n-type AlFe$_2$B$_2$ increases with the 6.9\% FeB impurity phase. Further, the Ga, and Ge addition decrease the Seebeck coefficient of Al$_{1.2}$Fe$_2$B$_2$ and Sb doping improves the S value. The effect of Ga and Ge doping on S can be attributed to the composite effect of additional impurity phases like AlB$_2$ and Al$_{13}$Fe$_4$. In Sb doping, the FeB impurity phase is very low; however, it has an additional AlSb secondary phase, which exhibits p-type conduction with S of around 300 $\mu$V/K and very low electrical conductivity. The average value of Seebeck coefficient obtained from the Gaussian fitting of the PSM data is -30 $\mu$V/K (Fig.~S3) which is greater than that of pristine Al$_{1.2}$Fe$_2$B$_2$ phase. The standard deviation (S.D.) for the Sb doped sample is higher and is due to the regions of positive Seebeck coefficient due to AlSb phase in the system.\\
\begin{figure}
    \centering
  \includegraphics[width=0.99\linewidth]{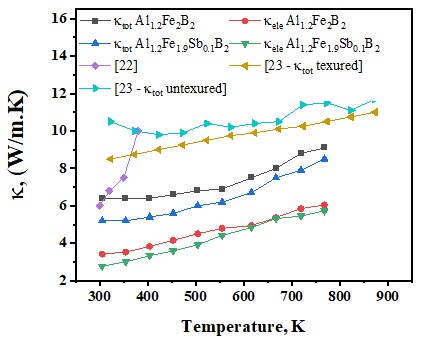}
  \caption{Thermal conductivities ($\kappa_{tot}$ and $\kappa_{ele}$) of Al$_{1.2}$Fe$_2$B$_2$ and Al$_{1.2}$Fe$_{1.9}$Sb$_{0.1}$B$_2$ compounds along with data reported in the literature for AlFe$_2$B$_2$.}
  \label{fig5a}
\end{figure}
The electrical conductivity ($\sigma$) of un-doped Al$_{1.2}$Fe$_2$B$_2$ is 0.46 ×10$^6$ S/m at room temperature and monotonously decreases to 0.35×10$^6$ S/m with an increase of temperature to 773\,K.  The $\sigma$ values at near-room temperature obtained are close to the values 0.38×10$^6$ S/m, and 0.50×10$^6$ S/m reported in the melt solidified and annealed samples, which decrease to 0.31×10$^6$ S/m and 0.375×10$^6$ S/m upon heating to 700\,K and 773\,K respectively.\cite{R5,R7} All the dopants Ag, Ni, Sb, Ge, and Ga addition decreases the Al$_{1.2}$Fe$_2$B$_2$ electrical conductivity at room temperature. With the increase in temperature, the conductivities of Sb, Ge, and Ga-doped compounds decrease, similar to bare Al$_{1.2}$Fe$_2$B$_2$; however, in Ag and Ni-doped compounds, the electrical conductivity decreases up to 550\,K and then increases, reaching 0.44×10$^6$ S/m at 773\,K comparable to RT value. The power factors of doped and undoped samples calculated from the S and $\sigma$ show that only Sb doping has some noticeable effect, nearly doubling it as compared to Al$_{1.2}$Fe$_2$B$_2$. Ga and Ge doping decrease the power factor compared to un-doped Al$_{1.2}$Fe$_2$B$_2$ and Ag; Ni has an insignificant effect.\\ 
The Al$_{1.2}$Fe$_{1.9}$Sb$_{0.01}$B$_2$ showing the highest power factor among the doped compounds and un-doped Al$_{1.2}$Fe$_2$B$_2$ were evaluated for thermal conductivity over the temperature range of 300 to 773\,K. Fig.\ref{fig5a} shows the variation of $\kappa_{tot}$ and $\kappa_{ele}$ of Al$_{1.2}$Fe$_2$B$_2$ and Al$_{1.2}$Fe$_{1.9}$Sb$_{0.1}$B$_2$ with temperatures. For comparison, the $\kappa_{tot}$ reported for AlFe$_2$B$_2$ prepared through the reactive hot pressing of the element powders and melt route is also included (data re-plotted from reference \cite{R6,R7}).\\
Unlike most thermoelectric materials, the thermal conductivity of AlFe$_2$B$_2$ increases with temperatures. The $\kappa_{tot}$ of Al$_{1.2}$Fe$_2$B$_2$ is 6.4 at 300\,K, increasing to 9.1 W/m.K at 773\,K. The $\kappa_{tot}$ of un-doped Al$_{1.2}$Fe$_2$B$_2$ obtained is lesser than the one reported by Kota et al. for un-textured and textured AlFe$_2$B$_2$ prepared by reactive hot-pressing of elemental powder followed by sinter forging.\cite{R7} However, the $\kappa_{tot}$ of 300\,K is comparable to 6.8 W/m.K at around 300\,K reported for melt and solidified AlFe$_2$B$_2$ sample\cite{R22,R24} The $\kappa_{ele}$ calculated from the measured electrical conductivity values using Wiedemann- Franz law $\kappa_{ele}$=L$_0\sigma T$, L$_o$ is Lorenz number = 2.44$\times$10$^{-8}$ W$.$K$^2$) suggests that the dominant contribution to $\kappa_{tot}$ comes from $\kappa_{ele}$. The difference in $\kappa_{tot}$ between this work and the Kota et al. could be attributed to the difference in the $\sigma$(T) and volume and physical characteristics of the secondary phase/s. The presence of secondary phases, FeB and Fe$_2$B could contribute to $\kappa_{tot}$ in this sample, considering the $\kappa$ of FeB is much greater than AlFe$_2$B$_2$.\cite{R13}\\
Doping with Sb decreases the thermal conductivity significantly. The presence of multiple submicron-size secondary phases like FeB and AlSb, apart from the noticeable extent of Sb dissolution in the AlFe$_2$B$_2$ lattice, appears to reduce the $\kappa_{tot}$ significantly. It is known that total thermal conductivity consists of electronic and phonon thermal conductivities.  From the electrical conductivity, it was found that both Al$_{1.2}$Fe$_2$B$_2$ and Al$_{1.2}$Fe$_{1.9}$Sb$_{0.1}$B$_2$ have almost similar behavior with temperature, contributing significantly to the overall $\kappa$. This further suggests that decreased $\kappa_{tot}$ in Sb doping mainly comes from reduced phonon thermal conductivity due to the scattering by secondary phase/s, resulting in lower thermal conductivity. The simultaneous effect of increased Seebeck coefficient and reduced thermal conductivity improves zT in the Sb-doped Al$_{1.2}$Fe$_2$B$_2$ sample. A maximum zT of 0.04 is achieved at 773 K for Al$_{1.2}$Fe$_{1.9}$Sb$_{0.1}$B$_2$.\\
\begin{figure}
    \centering
  \includegraphics[width=0.8\linewidth]{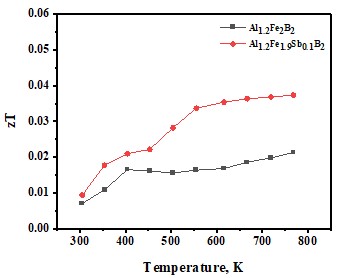}
  \caption{shows the zT variations as a temperature function for both samples.}
  \label{fig6a}
\end{figure}
\section{Conclusion}
The AlFe$_2$B$_2$ and AlFe$_{2-x}$M$_x$B$_2$ with M (Ag, Ni, Sb, Ga, and Ge) with 20\% Al above the stoichiometry were synthesized by melting, annealing, and powder consolidation.  The compound produced predominantly shows orthorhombic AlFe$_2$B$_2$ as the primary phase with single FeB in un-doped and multiple secondary phases in doped samples. The dopants Ag, Ni, and Sb added predominantly react with the Al, producing secondary phases like Ag$_5$Al, Al$_3$Ni$_2$, Al$_5$Ni$_3$, AlSb, and AlB$_2$. The structural parameters of the AlFe$_2$B$_2$ phase in these compounds are similar to the un-doped one. In the Ga and Ge doped compounds, the AlFe$_2$B$_2$ phase lattice volume increases mainly in the ‘$a$’ and ‘$c$’ lattice directions, suggesting the possible partial dissolution of these elements in the AlFe$_2$B$_2$ matrix, which is further confirmed by SEM-EDS analysis. The Ge-doped sample, unlike others, is free from FeB, instead containing AlB$_2$ as the secondary phase in the microstructure. A correlation between the formation of secondary phases in doped samples is further established from DSC-TG analysis. The thermoelectric characteristics viz Seebeck coefficient and electrical conductivity of the Ag, Ni, Ga, and Ge doped compounds are comparable to un-doped AlFe$_2$B$_2$. However, the Sb-doped sample exhibited a noticeable improvement in power factor and decreased thermal conductivity, resulting in a maximum zT of 0.04 at 773K.\\
\section*{Conflicts of interest}
There are no conflicts to declare.
\section*{Acknowledgements}
The authors are grateful to the Director, ARC-I for his support. This work was carried out with the financial support of the Department of Science and Technology, Government of India, through a research grant (No. AI/1/65/ARCI/2014). \\
\section{Supporting Information}
XRD pattern of Al$_{1.2}$Fe$_{2-x}$Sb$_x$B$_2$ ($x$=0.1, 0.25 and 0.5), DSC curve of Al$_{1.2}$Fe$_2$B$_2$ and Al$_{1.2}$Fe$_{1.9}$M$_{0.1}$B$_2$ (M=Sb, Ga, Ge), Spatial distribution of Seebeck coefficient ($S$) at room temperature for Al$_{1.2}$Fe$_2$B$_2$, and Sb doped Al$_{1.2}$Fe$_2$B$_2$.\\
%


%
%

\begin{figure*}
    \centering
  \includegraphics[width=0.8\linewidth]{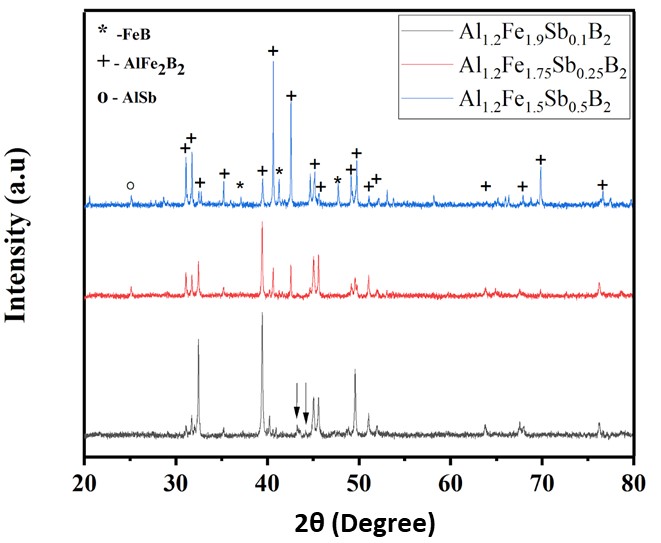}
  \caption{Fig.S1.XRD pattern of Al$_{1.2}$Fe$_{2-x}$Sb$_x$B$_2$ ($x$=0.1, 0.25 and 0.5). Arrow mark shows peaks corresponding to Al$_{13}$Fe$_4$ phase.}
  \label{figS1}
\end{figure*}

\begin{figure*}
    \centering
  \includegraphics[width=0.8\linewidth]{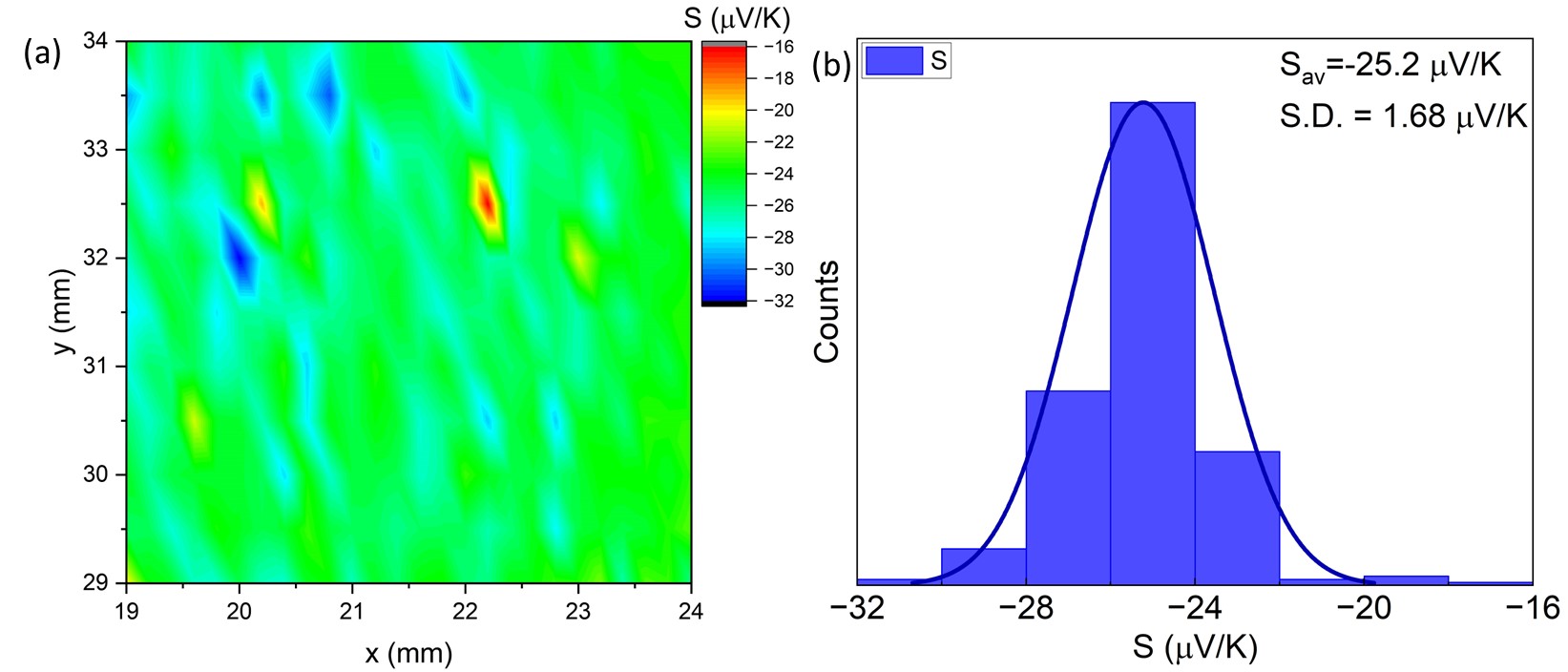}
  \caption{Fig.S2(a) Spatial distribution of Seebeck coefficient (S) at room temperature over 5mm × 5mm area in Al$_{1.2}$Fe$_2$B$_2$ samples measured using potential Seebeck microprobe. (b) The Seebeck coefficient obtained at the surface of the Al$_{1.2}$Fe$_2$B$_2$ sample is fitted with Gaussian function.}
  \label{figS2}
\end{figure*}

\begin{figure*}
    \centering
  \includegraphics[width=0.8\linewidth]{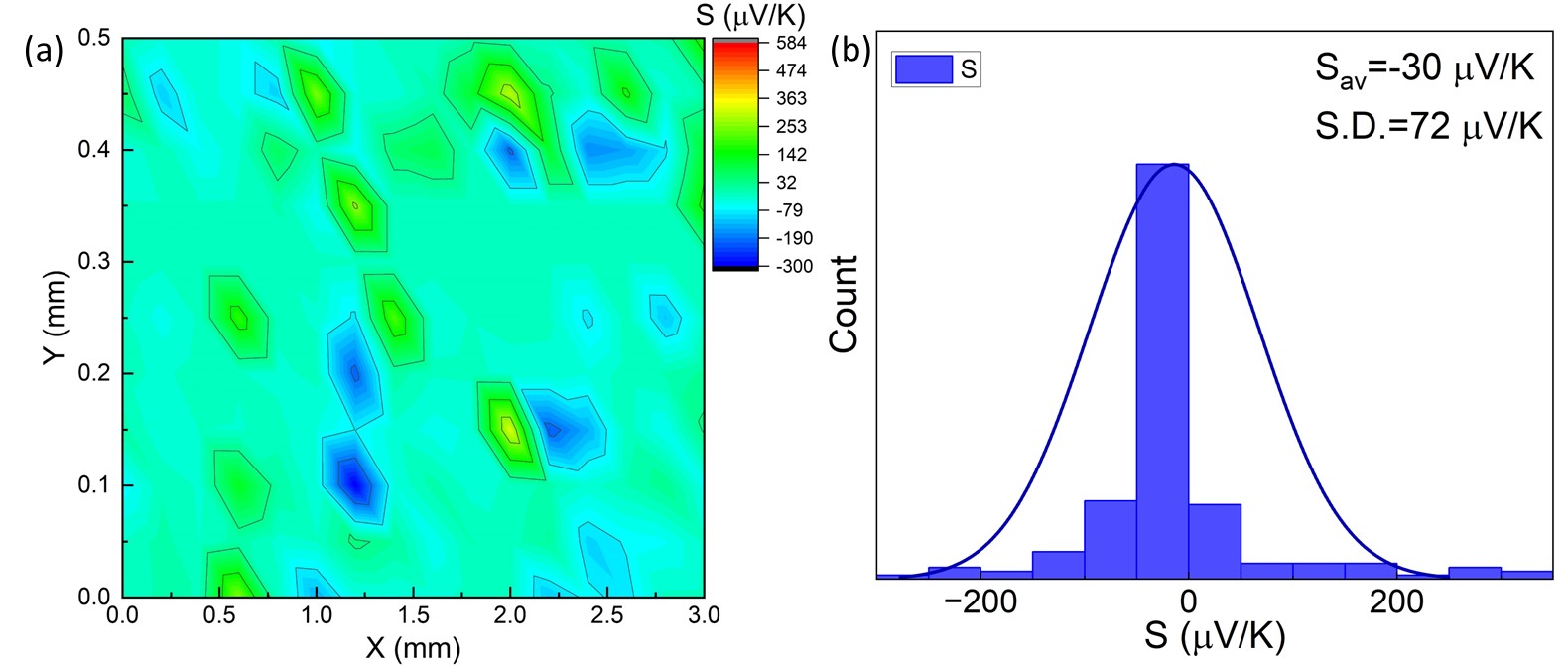}
  \caption{Fig.S3(a) Spatial distribution of Seebeck coefficient in Sb doped Al$_{1.2}$Fe$_2$B$_2$ sample and (b) corresponding distribution is fitted using a Gaussian function.}
  \label{figS3}
\end{figure*}

\end{document}